\documentclass[twocolumn,showpacs,aps,prl,superscriptaddress]{revtex4}

\usepackage{graphicx}
\usepackage{dcolumn}
\usepackage{amsmath}
\usepackage{epsfig}

\input babarsym

\newcommand{\BABARPubYear}    {11}
\newcommand{\BABARPubNumber}  {019}

\newcommand{\SLACPubNumber} {14541}

\begin{document}

\preprint{\babar-PUB-\BABARPubYear/\BABARPubNumber} 
\preprint{SLAC-PUB-\SLACPubNumber} 

\begin{flushleft}
%\babar\ Analysis Document \#2410, version 9\\
\babar-PUB-\BABARPubYear/\BABARPubNumber\\
SLAC-PUB-\SLACPubNumber\\[10mm]
%arXiv:\LANLNumber\ [hep-ex]\\[10mm]
\end{flushleft}

\title{
{\large \bf
Search for hadronic decays of a light Higgs boson in the radiative 
decay $\Upsilon \to \gamma A^0$
} 
}

% 
%% author list as of 01-Jul-2011 (387 authors)
%
\author{J.~P.~Lees}
\author{V.~Poireau}
\author{V.~Tisserand}
\affiliation{Laboratoire d'Annecy-le-Vieux de Physique des Particules (LAPP), Universit\'e de Savoie, CNRS/IN2P3,  F-74941 Annecy-Le-Vieux, France}
\author{J.~Garra~Tico}
\author{E.~Grauges}
\affiliation{Universitat de Barcelona, Facultat de Fisica, Departament ECM, E-08028 Barcelona, Spain }
\author{M.~Martinelli$^{ab}$}
\author{D.~A.~Milanes$^{a}$}
\author{A.~Palano$^{ab}$ }
\author{M.~Pappagallo$^{ab}$ }
\affiliation{INFN Sezione di Bari$^{a}$; Dipartimento di Fisica, Universit\`a di Bari$^{b}$, I-70126 Bari, Italy }
\author{G.~Eigen}
\author{B.~Stugu}
\affiliation{University of Bergen, Institute of Physics, N-5007 Bergen, Norway }
% ...
\collaboration{The \babar\ Collaboration}
\noaffiliation

\date{August 17, 2011}

\begin{abstract}
We search for hadronic decays of a light Higgs boson ($A^0$) produced in
radiative decays of an \TwoS\ or \ThreeS\ meson, 
$\Upsilon \to \gamma A^0$. 
The data have been recorded by the \babar\ experiment 
at the \ThreeS\ and \TwoS\ center of mass energies,
and include $(121.3 \pm 1.2)\times 10^6$ \ThreeS\ and 
$(98.3 \pm 0.9)\times 10^6$ \TwoS\ mesons. 
No significant signal is observed.
We set 90\% confidence level upper limits 
on the product branching fractions 
$\mathcal{B}(\Upsilon(nS) \to \gamma A^0) \cdot
\mathcal{B}(A^0 \to {\rm hadrons})$ ($n = 2$ or 3)
that range from $1 \times 10^{-6}$ for an $A^0$ mass of 0.3~\gevcc\ to 
$8 \times 10^{-5}$ at 7~\gevcc.  
\end{abstract}


\maketitle

A light CP-odd Higgs boson is expected in extensions to the 
Standard Model such as
non-minimal Supersymmetry \cite{ref:nmssm}. 
Light, in this context, means a mass less than that of the \OneS\ meson. 
Such a Higgs boson could be produced in radiative decays of
the $\Upsilon(nS)$ mesons \cite{ref:radups}, $\Upsilon(nS) \to \gamma A^0$, 
where in this analysis, $n = 2$ or 3.  
\babar\ has previously searched for this process where the 
$A^0$ decays to muons \cite{ref:mumu}, taus \cite{ref:tautau},
or invisibly \cite{ref:invisa,ref:invisb}.
CLEO has used its 
\OneS\ data sample to search in the muon pair and 
tau pair final states \cite{ref:cleo}. 
\babar\ has also searched for violations of lepton universality
in \OneS\ decay \cite{ref:universality}, which
could arise if the 
$A^0$ has the expected quantum numbers
$J^{PC} = 0^{-+}$ and mixes with 
the $\eta_b(1S)$ \cite{ref:domingouni}. 

Supersymmetry models in which $\tan^2\beta$ is not small predict that
the $A^0$ will decay predominantly into the heaviest kinematically available
down-type fermion pair. The earlier experimental results have ruled out
much of the 
parameter space \cite{ref:update, ref:domingoupdate}. 
Regions not excluded tend to be dominated by hadronic decays,
including decays to gluon pairs, $gg$, at smaller $\tan^2\beta$, 
and to charm quark pairs, $c \bar c$, at
higher $A^0$ mass.

This analysis searches for hadronic decays of the $A^0$ in the mass range
$2 m_\pi < m_{A^0} < 7$~\gevcc\ without attempting to specify the 
underlying partons to which the $A^0$ decays.  
The analysis nominally assumes that the $A^0$ is CP-odd, but also 
relaxes this assumption to obtain results without specifying the 
CP state.

The data were collected by the \babar\ detector~\cite{ref:detector}
at the \pep2\ asymmetric-energy \epem\ collider at the SLAC National
Accelerator Laboratory. 
They consist of 27.9~\invfb\ at 
the center of mass (c.m.) 
energy of the \ThreeS\ and 13.6~\invfb\ at the \TwoS, 
corresponding to $N_{3S} = (121.3 \pm 1.2)\times 10^6$ \ThreeS\ and
$N_{2S} = (98.3 \pm 0.9)\times 10^6$ \TwoS\ mesons. We also
use a continuum ({\em i.e.}, non-$\Upsilon(nS)$) background sample
consisting of 
78.3~\invfb\ of data collected 
at the c.m.\ energy of the 
\FourS, plus 11.8~\invfb\ of data recorded 30--40~\mev\ below the 
\TwoS, \ThreeS, or \FourS\ c.m.\ energies.
All of the data used here were recorded after the 
installation of an upgraded muon identification system~\cite{ref:LST}. 

Simulated signal events with various $A^0$ masses are used in the analysis.
The 
EvtGen event generator \cite{ref:evtgen} is used 
to simulate particle decays. The 
$A^0$ is simulated as a spin-0 particle, with equal branching fractions
to whichever of $g g$, $s \overline s$, and $c \overline c$ are 
kinematically available. 
Simulated events are produced both with and without the assumption that the 
$A^0$ is CP-odd.  
JETSET \cite{ref:jetset} is used to hadronize the partons, and 
\textsc{Geant4} \cite{ref:geant4} is used to simulate the detector response. 

The search for the $A^0$ uses hadronic 
events in which the full event energy is 
reconstructed. 
The selection criteria were optimized using simulated signal events 
and the continuum data set.  
The highest energy photon in each event
is taken to be the radiative photon from the 
$\Upsilon(nS)$ decay. 
The $A^0$ candidate is 
constructed by adding the four-momenta of 
the remaining particles
in the following order. The first added are 
$\KS \to \pi^+\pi^-$ candidates that 
have mass within 25~\mevcc\ of the 
true mass \cite{ref:pdg}, and whose reconstructed vertices are 
separated from the interaction point by at least three times the 
uncertainty on the vertex location. 
Charged hadron identification is then used to 
assign the proton, $K^\pm$, or $\pi^\pm$ mass to charged tracks. 
Tracks are labeled protons only 
if they are in the angular acceptance of the DIRC hadron 
identification system \cite{ref:detector}, 
and if there is another track identified as an anti-proton. 
Neutral pion candidates are formed from pairs of photons,
requiring the 
invariant mass of the photon pair to be 
between 100~\mevcc\ and 160~\mevcc, and 
to have a \piz\ energy
greater than 200~\mev. 
Finally, any remaining unused photons are added. 
Photons, including those used to reconstruct \piz\ mesons, 
are required to have a minimum energy of 90~MeV. All energies and
momenta are in the c.m.\ frame. 

Events are required to have a radiative photon energy 
greater than 2.5~\gev\ (\ThreeS) or 2.2~\gev\ (\TwoS) 
and to have at least two charged tracks among the $A^0$ decay products. 
The $A^0$ mass resolution is improved by constraining 
the radiative photon and all $A^0$ decay products 
to come from a common vertex, and the sum of the photon and $A^0$
four-momenta to be that of the c.m.\ system. 
To ensure that the full event energy is correctly reconstructed, the
probability of the $\chi^2$ of the constrained fit is required to be
greater than a value that ranges from 0 at low $A^0$ mass to 0.01 at 
$m_{A^0} = 7$~\gevcc. 
Events in which $m_{A^0} > 5$~\gevcc\ are 
rejected if the radiative photon, when combined with any other 
photon in the event, forms an invariant mass
within 50~\mevcc\ of the \piz\ mass,  
or, for $m_{A^0} > 6$~\gevcc, within 50~\mevcc\ of the $\eta$ mass. 

Additional criteria are used to reject
radiative Bhabha events,
$\epem \to \gamma \epem$, or radiative muon pairs, 
$\epem \to \gamma \mu^+ \mu^-$. 
An event is rejected if it was identified as a Bhabha at the
trigger level, if either of the 
two highest-momentum tracks is identified as an electron or 
a muon, or if the angle between
the radiative photon and the second-highest momentum track is 
less than 1 radian. These criteria reject 96\% of the 
continuum sample at a cost of 10--20\% in signal efficiency, and, 
according to simulation, reduce these backgrounds to negligible levels. 

The analysis proceeds along two parallel paths   
labeled ``CP-all'', in which 
no assumption is made on the CP nature of the $A^0$, and
``CP-odd'', in which it is assumed to be CP-odd. Events in 
which the $A^0$ decays to $\pi^+ \pi^-$ or 
$K^+ K^-$ are excluded from the CP-odd analysis.  

\begin{figure}
\includegraphics[width=\columnwidth]{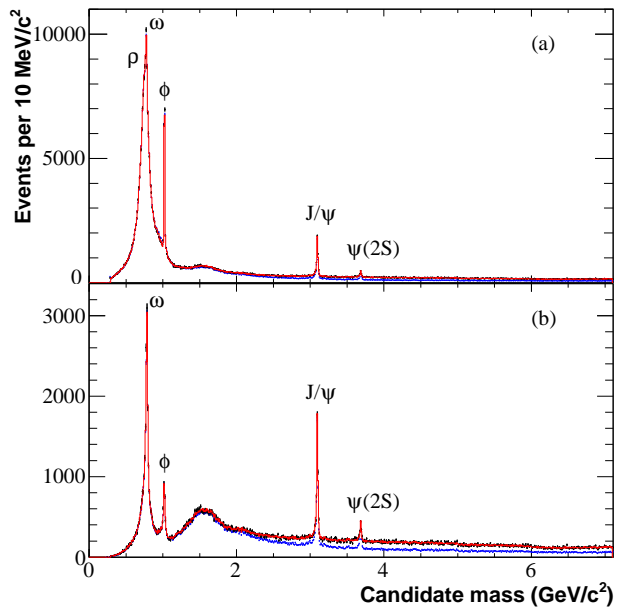}
\caption{
Candidate mass spectrum in the (a) CP-all and (b) CP-odd analyses. 
The top
curve in each plot is the on-peak data overlaid (in red) with the 
background fit described in the text, 
while the bottom curve (blue) 
is the scaled continuum data. The prominent initial state radiation 
resonances are labeled.    
}
\label{fig:massspectrum} 
\end{figure}

The analysis selects
371,740 events (CP-all) or 171,136 events (CP-odd) 
in the combined \TwoS\ and
\ThreeS\ (``on-peak'') data set with 
$0.29 < m_{A^0} < 7.1$~\gevcc\ (Fig.~\ref{fig:massspectrum}).

An $A^0$ signal would appear as a narrow peak in the candidate  mass spectrum. 
The number of signal events 
at a particular hypothesis mass is computed as the 
number of events in a
mass range (``window'') centered on that value, less the number 
of background events in the window. The width of the window 
depends on the $A^0$ mass resolution and was optimized along with the
other selection criteria.
It varies for CP-all from 
3 to 26~\mevcc\ as $m_{A^0}$ increases from 0.29 to 7~\gevcc. The 
CP-odd windows are the same width as CP-all above 2~\gevcc, 
but are larger at lower masses. 

Background events are from $\Upsilon(nS)$ decays and from continuum.
Continuum, which is 
dominant, mostly consists 
of the initial state radiation (ISR)
production of a light vector meson 
(clearly visible in Fig.~\ref{fig:massspectrum}) and non-resonant hadrons. 
The $\Upsilon(nS)$ backgrounds are 
primarily radiative decays to a light meson or non-resonant hadrons.
At the highest $A^0$ candidate masses, there is 
an additional contribution from hadronic $\Upsilon(nS)$ decays in which
a \piz\ daughter is misidentified as the radiative photon. 
Simulation indicates that the fraction of $\BB$ events satisfying the
selection criteria is negligible, so events recorded at the 
\FourS\ c.m.\ energy can be used in the continuum sample. 

The number of background events is obtained from a fit to the data that
contains three components: continuum, 
non-resonant $\Upsilon(nS)$ radiative decay, 
and resonant $\Upsilon(nS)$ radiative decay. 
The continuum component is the 
candidate mass spectrum of the 
continuum data set multiplied by a 
normalization factor $C_N$ ($\approx 0.5$). 
Because the efficiency for detecting the ISR photon depends on c.m.\ energy,
$C_N$ is not simply the ratio of integrated luminosities. 
It is left as a free parameter
in the nominal fit, but, as described below, is fixed to a 
calculated value for systematic studies. 
The non-resonant $\Upsilon(nS)$ component is a 16-knot cubic spline, 
fixed to 0 at the minimum $A^0$ mass. 
The resonant component includes five relativistic 
Breit-Wigner functions to represent the resonances for which 
CLEO saw some evidence
in the study of 
$\OneS \to \gamma h^+ h^-$ ($h = \pi$ or $K$) \cite{ref:cleotwobody}: 
$f_0(980)$, $f_2(1270)$, 
$f_2^\prime(1525)$, $f_0(1710)$, and $f_4(2050)$. 
The masses and widths are fixed \cite{ref:pdg} and 
possible interference between the resonances is neglected in the fit.  
These resonances are all broad compared to an $A^0$ signal.
The spacing of the knots, typically 0.5~\gevcc, is large enough that
the cubic spline cannot conform to a narrow resonance. 

\begin{figure}
\includegraphics[width=\columnwidth]{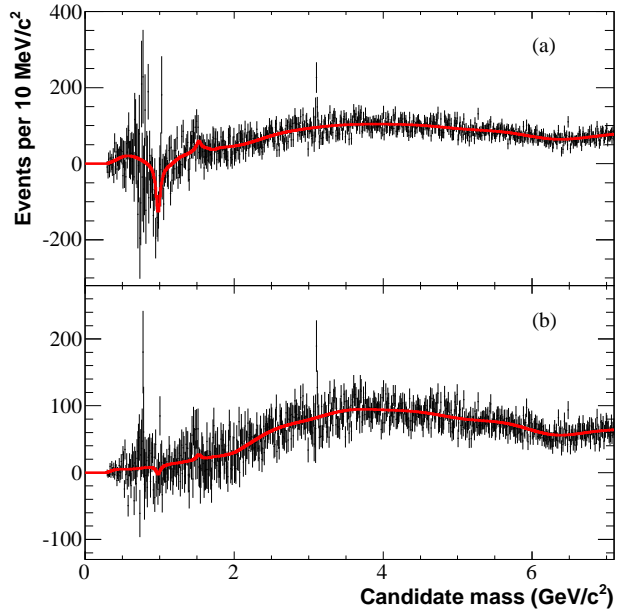}
\caption{
$A^0$ candidate mass spectrum after continuum subtraction, overlaid
with fit. (a) CP-all analysis; (b)
CP-odd analysis. 
}
\label{fig:upsspectrum} 
\end{figure}

The background fit (Fig.~\ref{fig:massspectrum}) has 21 free parameters, 
and is made to  1362 bins of width 
5~\mevcc, ranging from 0.29 to 7.1~\gevcc. 
The fit $\chi^2$ are 1268 (CP-all) and 1293 (CP-odd) for 1341 degrees
of freedom. 
Subtracting the normalized 
continuum mass spectrum 
from both the data and the fit gives the $\Upsilon(nS)$ decay 
spectrum and the 
non-resonant and resonant radiative $\Upsilon$ decay components
of the fit (Fig.~\ref{fig:upsspectrum}).

The uncertainty on the background in each mass window
is both statistical and systematic. 
The systematic error is the sum in quadrature of the change in the 
total background arising from each of 17 alternative fits: the five
nominal light resonances are removed one at a time, and eleven additional
resonances are included one at a time. The eleven are 
established resonances \cite{ref:pdg} with even total angular momentum, 
charge conjugation quantum number of $+1$, and Isospin 0.  
The seventeenth alternative fit is performed with
$C_N$ fixed to 
the mid-point of the range of values found from four different methods
of determining it. Two of the methods are the nominal fits to the 
CP-odd and CP-all samples, and two use 
ISR-produced narrow resonances in four different final states:  
$\epem \to \gamma \omega$, $\omega \to \pi^+ \pi^- \pi^0$; 
$\epem \to \gamma \phi$, $\phi \to K^+ K^-$; 
$\epem \to \gamma \jpsi$, $\jpsi \to \ge 4$ charged tracks, with 
no \piz; and
$\epem \to \gamma \jpsi$, $\jpsi \to \ge 4$ charged tracks, with 
one \piz. 
First, the number of each of these resonances is compared in 
on-peak and continuum data.  Second, the same ratios are 
obtained  
using simulated samples of these
ISR events, together with the calculated production 
cross sections \cite{ref:isr} and the recorded luminosities. 
The resulting value of $C_N$ is 4.5\% larger than nominal for CP-all, and
2.7\% for CP-odd. 
The fit qualities are good in all alternative fits. 
The systematic errors
are small compared to statistical errors except near resonances. 

\begin{figure}
\includegraphics[width=\columnwidth]{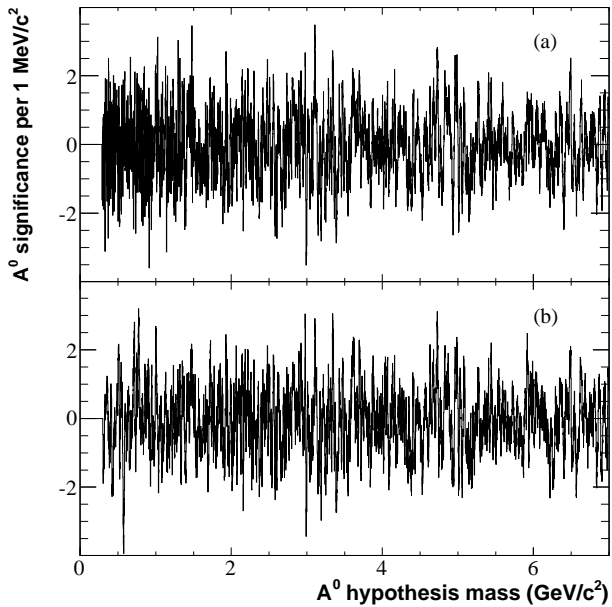}
\caption{
Statistical significance (events divided by statistical error) of the
$A^0$ signal as a function of mass, for 
(a) CP-all analysis, and (b) CP-odd analysis.
}
\label{fig:higgssig}
\end{figure}

\begin{figure}
\includegraphics[width=\columnwidth]{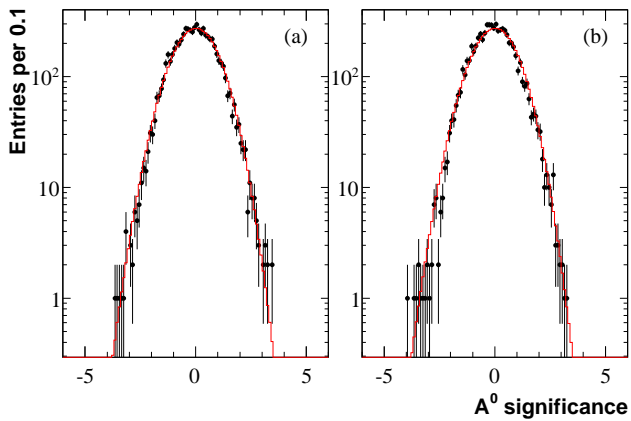}
\caption{
Histogram of the statistical significance of the 
$A^0$ signal for (a) the 6710 masses considered in the
CP-all analysis, and for (b) the 6701 masses in the CP-odd analysis.
The overlaid curve shows the distribution expected in the 
absence of signal. 
}
\label{fig:higgsdev}
\end{figure}

The $A^0$ signal is evaluated at hypothesis masses that range from 
0.291~\gevcc\ to 7.000~\gevcc\ in 
1~\mevcc\ steps for the CP-all analysis (6710 mass hypotheses), 
and from 0.300~\gevcc\ to 7.000~\gevcc\ in 1~\mevcc\ steps
for CP-odd (6701 masses). 
Figure~\ref{fig:higgssig} shows the nominal statistical 
significance of the resulting $A^0$ signal, 
defined as the number of events divided by 
the statistical error, as a function
of mass. 
The largest upwards fluctuations
are $3.5\sigma$ at 
3.107~\gevcc\ for CP-all 
and $3.2\sigma$ at 0.772~\gevcc\ for CP-odd. Including background 
systematic errors, the significance of these two, which are located 
near the \jpsi\ and $\rho$ resonances respectively, are reduced to 
$2.8\sigma$ and $2.2\sigma$. 
The largest remaining fluctuations are 
$2.9\sigma$ at 
1.295~\gevcc\ for CP-all 
and $3.1\sigma$ at 4.727~\gevcc\ for CP-odd.
Figure~\ref{fig:higgsdev}
histograms the statistical significance of the signal measured at each 
mass, overlaid with the distribution
expected in the absence of a signal. 

The signal extraction technique is studied using many 
simulated experiments. 
Each experiment consists
of two candidate mass distributions, 
one for on-peak data, and the other for continuum. 
The continuum 
event distributions are obtained from the full 
\FourS\ data, 
which is 11 times larger than the on-peak data set. 
The non-resonant $\Upsilon(nS)$ events are 
generated from a smooth threshold curve, and the resonant events 
are generated from relativistic Breit-Wigner functions. The full signal 
extraction is then performed. 
The average bias
on the $A^0$ signal yield is less than 1.5 events for all masses when there is
no signal. 

These studies are also used to calculate the expected distribution of
statistical significance in the absence of signal (Fig.~\ref{fig:higgsdev})
and to evaluate 
the significance of the 
largest apparent $A^0$ signals. 
The fraction of 
background-only CP-all simulated experiments
that contain a fluctuation of nominal statistical significance 
$\ge 3.5\sigma$ is 33\%. The fraction 
of CP-odd simulated experiments that contain a fluctuation
$\ge 3.2\sigma$ is 63\%. 
We therefore see no evidence of signal. 
The studies further indicate that 
large correlations between the resonant and non-resonant $\Upsilon(nS)$ 
components make the uncertainties on the yields of the
resonances unreliable. 

In the absence of a significant signal, we calculate a 90\% 
confidence level (CL) upper limit for each hypothesis mass 
on the product branching 
fractions
$\mathcal{B}_{3S} \equiv 
\mathcal{B}(\ThreeS \to \gamma A^0) \cdot
\mathcal{B}(A^0 \to {\rm hadrons})$ and 
$\mathcal{B}_{2S} \equiv 
\mathcal{B}(\TwoS \to \gamma A^0) \cdot
\mathcal{B}(A^0 \to {\rm hadrons})$, assuming that 
the \ThreeS\ and \TwoS\ decays are described by the same 
matrix element. This implies that $\mathcal{B}_{2S} = 
\mathcal{B}_{3S} \cdot \Gamma_{3S} / \Gamma_{2S}\cdot R(m_{A^0})$, where
$\Gamma_{3S}$ and $\Gamma_{2S}$ are the full widths of the 
\ThreeS\ and \TwoS\ respectively, and $R$ accounts for the 
difference in phase space. $R$ is within a few percent of 
unity for all $A^0$ masses. 

The calculation uses the 
relationship 
$\hat{N} = \hat{B} + N^\prime_{3S} \cdot \mathcal{B}_{3S} \cdot \epsilon$,
where $\hat{N}$ is the expected number of observed events
for the given value of $\mathcal{B}_{3S}$, 
$\hat B$ is the expected background, 
$N^\prime_{3S} \equiv N_{3S} + N_{2S}\cdot \Gamma_{3S} / \Gamma_{2S}\cdot R(m_{A^0})$, 
and $\epsilon$ is the signal efficiency. 
We calculate a likelihood $\mathcal{L}(\mathcal{B}_{3S})$, 
defined as the probability of observing $N$ or fewer 
events given that value of $\mathcal{B}_{3S}$, where $N$ is the number 
actually observed. 
$\mathcal{L}(\mathcal{B}_{3S})$ is obtained by 
integrating over the uncertainties 
in $\hat B$, $N_{2S}$, $N_{3S}$, and $\epsilon$, which are assumed
to be Gaussian.  
The 90\% CL upper limit 
$\mathcal{B}_{90}$ is calculated assuming a uniform prior above 0:
$\int_0^{\mathcal{B}_{90}} \mathcal{L}(\mathcal{B}_{3S}) d\mathcal{B}_{3S} = 0.90 \int_0^{\infty} \mathcal{L}(\mathcal{B}_{3S}) d\mathcal{B}_{3S}$.

The efficiency is calculated using simulated events.
The efficiency for the CP-all analysis ranges from a peak of 22\% 
near $m_{A^0} = 0.6$~\gevcc\ to less than 1\% at high masses, while for the 
CP-odd analysis it ranges from 12\% near 0.9~\gevcc\ to less than
1\% at high masses.

\begin{figure}
\includegraphics[width=\columnwidth]{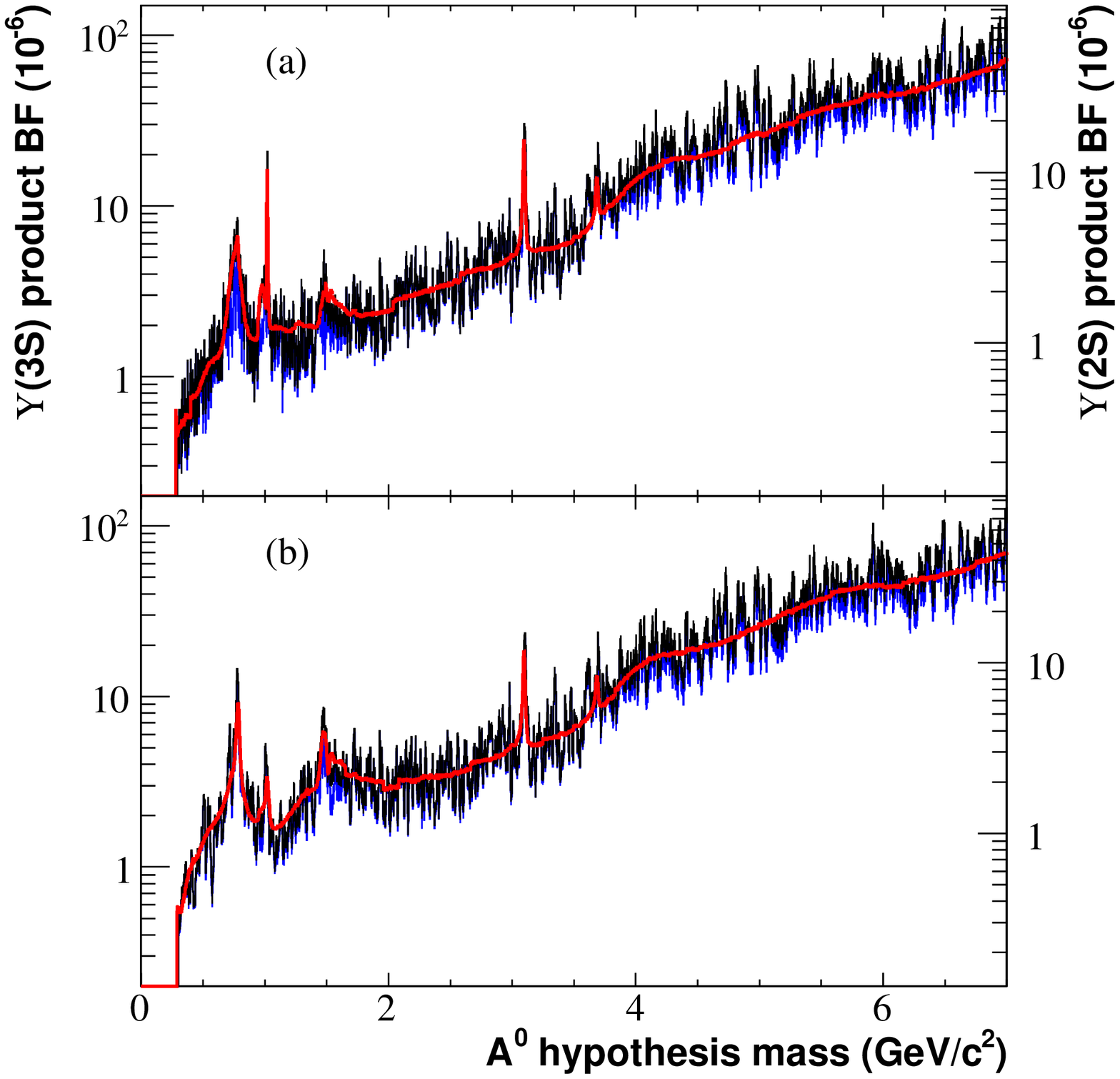}
\caption{
90\% CL 
upper limits on product branching fractions (BF) 
(left axis) $\mathcal{B}(\ThreeS \to \gamma A^0) \cdot
\mathcal{B}(A^0 \to {\rm hadrons})$ and (right axis)  
$\mathcal{B}(\TwoS \to \gamma A^0) \cdot
\mathcal{B}(A^0 \to {\rm hadrons})$, for 
(a) CP-all
analysis, and (b) CP-odd analysis. The overlaid curves in red are the
limits expected from simulated experiments, while the 
blue curves are the limits from statistical 
errors only. The \TwoS\ limits 
do not include the phase space factor, which is at most
a 3.5\% correction.  
}
\label{fig:UL}
\end{figure}

The uncertainty on the efficiency is typically 11\% (CP-all) 
or 7\% (CP-odd)  below the 
$c \overline c$ threshold, and 25\% above. 
This includes contributions
from uncertainty in tracking (1.5--3.5\% depending on mass), 
photon and \piz\ reconstruction (5--10\%), and particle 
identification (3--5\%), but the dominant contribution is due to 
the $A^0$ decay branching fractions.
This uncertainty is evaluated by varying the assumed branching 
fractions from 
50\% $s \overline s$ and 50\% $g g$ to 100\% $g g$
below the $c \overline c$ threshold, and from 
one-third each $g g$, $s \overline s$, and $c \overline c$
to  50\% $c \overline c$
and 25\% each of $g g$ and $s \overline s$ above the 
$c \overline c$ threshold.
The resulting systematic errors are 
8\% for CP-all or 4\% for CP-odd below the 
$c \overline c$ threshold, and 21\% above.
The resulting 90\% CL upper limits are shown in Fig.~\ref{fig:UL}. 

In conclusion, we have searched for 
hadronic final states of a light Higgs boson 
produced in radiative decays of the \TwoS\ or  \ThreeS\ and 
find no evidence of a signal.  
Upper limits on the product branching fraction
$\mathcal{B}(\Upsilon(nS) \to \gamma A^0) \cdot 
\mathcal{B}(A^0 \to {\rm hadrons})$ 
range from $1 \times 10^{-6}$ at 0.3~\gevcc\ to 
$8\times 10^{-5}$ at 7~\gevcc\ at the 90\% CL. 

We are grateful for the excellent luminosity and machine conditions
provided by our \pep2\ colleagues, 
and for the substantial dedicated effort from
the computing organizations that support \babar.
The collaborating institutions wish to thank 
SLAC for its support and kind hospitality. 
This work is supported by
DOE
and NSF (USA),
NSERC (Canada),
IHEP (China),
CEA and
CNRS-IN2P3
(France),
BMBF and DFG
(Germany),
INFN (Italy),
FOM (The Netherlands),
NFR (Norway),
MIST (Russia), and
PPARC (United Kingdom). 
Individuals have received support from CONACyT (Mexico), A.~P.~Sloan Foundation, 
Research Corporation,
and Alexander von Humboldt Foundation.

\end{document}